\documentstyle[aps,eqsecnum,preprint,floats,epsf,psfig]{revtex}
\renewcommand{\baselinestretch}{1.05}

\begin{document}
\def\be{\begin{eqnarray}}
\def\en{\end{eqnarray}}
\def\up{\uparrow}
\def\dw{\downarrow}
\def\non{\nonumber}
\def\la{\langle}
\def\ra{\rangle}
\def\etapp{{\eta^{(')}}}
\def\ov{\overline}
\def\vp{\varepsilon}
\def\half{{{1\over 2}}}
\def\a{{\cal A}}
\def\b{{\cal B}}
\def\c{{\cal C}}
\def\d{{\cal D}}
\def\pr{{\sl Phys. Rev.}~}
\def\prl{{\sl Phys. Rev. Lett.}~}
\def\pl{{\sl Phys. Lett.}~}
\def\np{{\sl Nucl. Phys.}~}
\def\zp{{\sl Z. Phys.}~}

\font\el=cmbx10 scaled \magstep2
{\obeylines
\hfill }

\vskip 1.5 cm

\centerline{\large\bf Exclusive Hadronic $D$ Decays to $\eta'$ and $\eta$}
\medskip
\bigskip
\medskip
\centerline{\bf Hai-Yang Cheng}
\medskip
\centerline{ Institute of Physics, Academia Sinica}
\centerline{Taipei, Taiwan 115, Republic of China}
\medskip
\centerline{\bf B. Tseng}
\medskip
\centerline{Department of Physics, National Cheng Kung University}
\centerline{Tainan, Taiwan 700, Republic of China}
\bigskip
\bigskip
\bigskip
\centerline{\bf Abstract}
\bigskip
{\small   
Hadronic decay modes $D^0\to(\ov K^0,\ov K^{*0})(\eta,\eta')$ and 
$(D^+,D_s^+)\to(\pi^+,\rho^+)(\eta,\eta')$ are studied in the generalized 
factorization approach.
Form factors for $(D,D_s^+)\to(\eta,\eta')$ transitions are carefully 
evaluated by taking into account the wave function normalization of the 
$\eta$ and $\eta'$. The predicted branching ratios are generally 
in agreement with experiment except for $D^0\to\ov K^0\eta',~D^+\to\pi^+
\eta$ and $D_s^+\to\rho^+\eta'$; the calculated decay rates for the 
first two decay modes are too small by an order of magnitude.
We show that the weak decays $D^0\to K^-\pi^+$ and $D^+\to K^+\ov 
K^0$ followed by resonance-induced final-state interactions 
(FSI), which are amenable technically,
are able to enhance the branching ratios of $D^0\to\ov K^0\eta'$ and 
$D^+\to\pi^+\eta$ dramatically without affecting the agreement 
between theory and experiment for $D^0\to\ov K^0\eta$ and 
$D^+\to\pi^+\eta'$. We argue that it is difficult to understand the 
observed large decay rates of $D_s^+\to \rho^+\eta'$ and $\rho^+\eta$
simultaneously; FSI, $W$-annihilation 
and the production of excess $\eta'$ from gluons
are not helpful in this regard. The large discrepancy between the 
factorization hypothesis and experiment for the ratio of $D_s^+\to\rho^+
\eta'$ and $D_s^+\to\eta' e^+\nu$ remains as an enigma.

}

\pagebreak
 
\section{Introduction}
The exclusive rare $B$ decays to the 
$\eta'$ have recently received a great deal of attention since the observed
large branching ratio of $B^-\to\eta' K^-$ by CLEO is substantially higher 
than the naive theoretical estimates (for a review, see \cite{CT98}).
It has stimulated many theoretical studies and speculation. It is natural
to reexamine the hadronic decays of the charmed mesons into the final
states containing an $\eta$ or $\eta'$. Experimentally, CLEO has recently 
remeasured the decay modes $(D_s^+,D^+)\to(\pi^+,\rho^+)\etapp$ 
\cite{CLEO}. Combined with the previous measurements of $D^0\to \ov K^{0(*)}
\etapp$, we see an $\eta'$ enhancement for $(D_s^+,D^+)\to\pi^+\eta'$ over 
$(D_s^+,D^+)\to\pi^+\eta$ and for $D^0\to\ov K^0\eta'$ over
$D^0\to\ov K^0\eta$ (see Table I). Also, very large branching ratios for
$D_s^+\to \rho^+\eta'$ and $D_s^+\to \rho^+\eta$ are confirmed by the new 
data \cite{CLEO}. Theoretically, the factorization approach of Bauer, Stech
and Wirbel (BSW) \cite{BSW87} predicts less $\eta'$ production than $\eta$ 
in $D^0\to\ov K^0\etapp$ and $D_s^+\to\pi^+\etapp$ decays, in disagreement
with experiment (see Table I). Moreover, the decay $D^+\to\pi^+\eta$ is 
severely suppressed in the BSW approach, about two orders of 
magnitude smaller than the experimental measurement. Likewise, the 
predicted branching ratios 
for $D^0\to\ov K^{*0}\eta$ and $D_s^+\to\rho^+\eta'$ are also too small.
Many different theoretical efforts have been made in the past
to explain the data \cite{Verma,Kamal,Lipkin92,Buccella}.

Care must be taken when applying the BSW form factors for $(D,D_s)\to(\eta,
\eta')$ transitions as the wave function normalizations of the $\eta$
and $\eta'$ are not taken into account in the original BSW analysis
\cite{BSW}.
In this paper we will
evaluate these form factors in a consistent way and present an updated
analysis in the generalized factorization approach. Then we proceed to
propose that final-state interactions (FSI) in resonance formation are 
responsible for the discrepancy between theory and experiment for 
the above-mentioned 
$\eta'/\eta$ ratios and for the decay rate of $D^+\to\pi^+\eta$. Since some
resonances are known to exist in the charm mass region and since the 
charm decay is not very energetic, FSI are expected to play an essential 
role in the nonleptonic charm decays. We shall show in the present paper that
$D^+\to\pi^+\eta$ and $D^0\to\ov K^0\eta'$ are essentially generated from
FSI. Finally, we will comment on the observed large branching ratio for
the decay $D_s^+\to\rho^+\eta'$.

\section{Generalized Factorization}
The effective weak 
Hamiltonian for nonleptonic charm decay relevant to the present paper is
\be
{\cal H}_{\rm eff} &=& {G_F\over\sqrt{2}}\Bigg\{ V_{cs}^*V_{ud}\Big(
c_1(\mu)(\bar ud)(\bar sc)+c_2(\mu)(\bar uc)(\bar sd)\Big)  \non \\   
&+& \sum_{q=d,s}V_{cq}^*V_{uq}\Big(c_1(\mu)O_1^q(\mu)+c_2(\mu)O_2^q(\mu)
\Big)\Bigg\},
\en 
with $O_1^q=\,(\bar uq)(\bar qc)$ and $O_2^q=\,(\bar uc)(\bar qq)$,
where $(\bar q_1q_2)\equiv \bar q_1\gamma_\mu(1-\gamma_5)q_2$ and 
$c_{1,2}(\mu)$ are the Wilson coefficient functions. 
The mesonic matrix elements of four-quark operators are customarily 
evaluated under the factorization approximation.
It is known that naive factorization fails to describe color-suppressed 
charm decays. Therefore, it is necessary and mandatory
to take into account nonfactorizable contributions to the weak decay 
amplitudes. For $D\to PP,\,VP$ decays ($P$: pseudoscalar meson, $V$: vector
meson), the effects of nonfactorization characterized by the 
parameters $\chi_1$ and $\chi_2$ can be lumped into the effective 
parameters $a_1$ and $a_2$ \cite{Cheng94}:
\be
a_1=c_1+c_2\left({1\over N_c}+\chi_1\right), \qquad a_2=c_2+c_1\left({1\over 
N_c}+\chi_2\right),
\en
where $N_c$ is the number of colors. If $\chi_{1,2}$ are universal (i.e. 
process independent) in charm decays, then we still have a new 
factorization scheme in which the decay amplitude is expressed in 
terms of factorizable contributions multiplied by the universal effective
parameters $a_{1,2}$. By treating $a_{1,2}$ as free parameters, they 
can be determined from experiment. For example, neglecting the 
$W$-exchange contribution and assuming that final-state interactions 
can be described by isospin phase shifts, we find that
\be \label{a1a2}
a_1(D\to \ov K\pi)=1.25, \qquad a_2(D\to\ov K\pi)=-0.51
\en
from the data of $D^0\to K^-\pi^+,~\ov K^0\pi^0$ and $D^+\to\ov K^0\pi^+$ 
decays. 

   We next consider the two-body decays of charmed mesons into the $\eta$
or $\eta'$. Neglecting $W$-exchange or $W$-annihilation, it is
easily seen that $D_s^+\to (\pi^+,\rho^+)\etapp$ proceed through the 
color-allowed external $W$-emission, $D^0\to\ov K^{0(*)}\etapp$ via 
the color-suppressed internal $W$-emission, and $D^+\to (\pi^+,\rho^+)\etapp$
receive contributions from both external and internal $W$-emission 
diagrams. Under the generalized factorization hypothesis, it is 
straightforward to write down the decay amplitudes of the charmed 
meson decays to the final state containing an $\eta$ or $\eta'$:
\be  \label{amp}
A(D^0\to\ov K^0\etapp) &=& {G_F\over\sqrt{2}} V_{cs}^*V_{ud}\,a_2\left(
X^{(D\etapp,K)}+2X^{(D,K\etapp)}\right),  \non \\
A(D^0\to\ov K^{*0}\etapp) &=& {G_F\over\sqrt{2}} V_{cs}^*V_{ud}\,a_2\left(
X^{(D\etapp,K^*)}+2X^{(D,K^*\etapp)}\right),  \non \\
A(D^+\to\pi^+\etapp) &=& {G_F\over\sqrt{2}}V_{cd}^*V_{ud}\left[ a_1X^{(D
\etapp,\pi)}+a_2\left(X_d^{(D\pi,\etapp)}-X_s^{(D\pi,\etapp)}\right)
+2a_1X^{(D,\etapp\pi)}\right],   \non \\
A(D^+\to\rho^+\etapp) &=& {G_F\over\sqrt{2}}V_{cd}^*V_{ud}\left[ a_1X^{(D
\etapp,\rho)}+a_2
\left(X_d^{(D\rho,\etapp)}-X_s^{(D\rho,\etapp)}\right)+2a_1X^{(D,\etapp\rho)}
\right],   \non \\
A(D_s^+\to\pi^+\etapp) &=& {G_F\over\sqrt{2}}V_{cs}^*V_{ud}\,a_1\left(X^{
(D_s\etapp,\pi)}+2X^{(D_s,\etapp\pi)}\right),  \non \\
A(D_s^+\to\rho^+\etapp) &=& {G_F\over\sqrt{2}}V_{cs}^*V_{ud}\,a_1\left(X^{
(D_s\etapp,\rho)}+2X^{(D_s,\etapp\rho)}\right),  
\en
where use of the approximation $V_{cs}^*V_{us}\approx -V_{cd}^*V_{ud}$ has 
been 
made and $X^{(DP_1,P_2)}$ denotes the factorizable amplitude with the meson
$P_2$ being factored out:
\be
X^{(DP_1,P_2)}=\la P_2|(\bar q_1 q_2)|0\ra\la P_1|(\bar q_3 c)|D\ra.
\en
Explicitly,
\be
X^{(D\etapp,K)} &=& if_K(m_D^2-m^2_\etapp)F_0^{D\etapp}(m^2_K),  \non \\
X^{(D\etapp,\pi)} &=& if_\pi(m_D^2-m^2_\etapp)F_0^{D\etapp}(m^2_\pi),  \non \\
X_q^{(D\pi,\etapp)} &=& if^q_\etapp(m_D^2-m^2_\pi)F_0^{D\pi^\pm}(m^2_\etapp),  
\non \\
X^{(D_s\etapp,\pi)} &=& if_\pi(m_{D_s}^2-m^2_\etapp)F_0^{D_s\etapp}(m^2_\pi),
\non \\
X^{(D\etapp,K^*)} &=& -2f_{K^*}m_{K^*}F_1^{D\etapp}(m^2_{K^*})(\vp\cdot 
p_{_D}),  \non \\
X^{(D\etapp,\rho)} &=& -2f_{\rho}m_{\rho}F_1^{D\etapp}(m^2_{\rho})(\vp\cdot 
p_{_D}),  \non \\
X^{(D_s\etapp,\rho)} &=& -2f_\rho m_\rho F_1^{D_s\etapp}(m^2_\rho)(\vp\cdot 
p_{_{D_s}}),  \non \\
X_q^{(D\rho,\etapp)} &=& -2f^q_\etapp m_\rho A_0^{D\rho}(m^2_\etapp)
(\vp\cdot p_{_D}),
\en
where $\la 0|\bar q\gamma_\mu\gamma_5 q|\etapp
\ra=if^q_\etapp p_\mu$, and form factors $F_0,~F_1$ and $A_0$ are those 
defined in \cite{BSW}. The amplitude $X^{(D,\etapp P)}$ in Eq.~(\ref{amp})
denotes $W$-exchange or $W$-annihilation, for example,
$X^{(D,\eta\pi)}=\la\eta\pi^+|(\bar ud)|0\ra\la 0|(\bar dc)|D^+\ra$. 

   To determine the decay constant $f_{\eta'}^q$, we need to know the
wave functions of the physical $\eta'$ and $\eta$ states which are related to
that of the SU(3) singlet state $\eta_0$ and octet state $\eta_8$ by
\be
\eta'=\eta_8\sin\theta+\eta_0\cos\theta, \qquad \eta=\eta_8\cos\theta-\eta_0
\sin\theta,
\en
with $\theta\approx -20^\circ$. When the $\eta-\eta'$ mixing angle is 
$-19.5^\circ$, 
the $\eta'$ and $\eta$ wave functions have simple expressions \cite{Chau1}:
\be  \label{etapp}
|\eta'\ra={1\over\sqrt{6}}|\bar uu+\bar dd+2\bar ss\ra, \qquad
|\eta\ra={1\over\sqrt{3}}|\bar uu+\bar dd-\bar ss\ra,
\en
recalling that
\be
|\eta_0\ra={1\over\sqrt{3}}|\bar uu+\bar dd+\bar ss\ra, \qquad
|\eta_8\ra={1\over\sqrt{6}}|\bar uu+\bar dd-2\bar ss\ra.
\en
At this specific mixing angle, $f_{\eta'}^u={1\over 2}f_{\eta'}^s$ in the
SU(3) limit. Introducing the decay constants $f_8$ and $f_0$ by 
$\la 0|A_\mu^0|\eta_0\ra=if_0 p_\mu$ and $\la 0|A_\mu^8|\eta_8\ra=if_8 
p_\mu$,
then $f_{\eta'}^u$ and $f_{\eta'}^s$ are related to $f_8$ and $f_0$ by
\be
f_{\eta'}^u={f_8\over\sqrt{6}}\sin\theta+{f_0\over\sqrt{3}}\cos\theta,
\qquad f_{\eta'}^s=-2{f_8\over\sqrt{6}}\sin\theta+{f_0\over\sqrt{3}}\cos
\theta.
\en
Likewise, for the $\eta$ meson
\be
f_{\eta}^u={f_8\over\sqrt{6}}\cos\theta-{f_0\over\sqrt{3}}\sin\theta,
\qquad f_{\eta}^s=-2{f_8\over\sqrt{6}}\cos\theta-{f_0\over\sqrt{3}}\sin\theta.
\en
Applying the results
\be
{f_8\over f_\pi}=1.38\pm 0.22, \qquad {f_0\over f_\pi}=1.06\pm 0.03,
\qquad \theta=-22.0^\circ\pm 3.3^\circ,
\en
extracted from a recent analysis of the data of $\eta,\eta'\to \gamma\gamma$
and $\eta,\eta'\to\pi\pi\gamma$ \cite{Holstein} yields
\be  \label{feta}
f_\eta^u=99\,{\rm MeV}, \quad f_\eta^s=-108\,{\rm MeV}, \quad
f_{\eta'}^u=47\,{\rm MeV}, \quad f_{\eta'}^s=131\,{\rm MeV}.
\en
\vskip 0.1cm

   To compute the form factors $F_0^{D\etapp}$ and $F_0^{D_s\etapp}$,
we will first apply the nonet symmetry relations
\be
\sqrt{6}F_0^{D\eta_8}(0)= \sqrt{3}F_0^{D\eta_0}(0)=F_0^{D\pi^\pm}(0),  
\non \\
-{\sqrt{6}\over 2}F_0^{D_s\eta_8}(0)= \sqrt{3}F_0^{D_s\eta_0}(0)=
F_0^{D_sK}(0),  
\en
to determine $F_0^{D\eta_{0,8}}(0)$ and $F_0^{D_s\eta_{0,8}}(0)$, and 
then relate them to the form factors $F_0^{D\etapp}$ and $F_0^{D_s\etapp}$
via
\be
F_0^{D\eta}= F_0^{D\eta_8}\cos\theta-F_0^{D\eta_0}\sin\theta, \qquad
F_0^{D\eta'} = F_0^{D\eta_8}\sin\theta+F_0^{D\eta_0}\cos\theta.  
\en
Using $F_0^{D\pi^\pm}(0)\approx F_0^{DK}(0)\approx 0.75$ as inferred from 
experiment \cite{Dpi}, and taking
$F_0^{D_s K}(0)\approx 0.76$ extracted from the data of $D_s^+\to K^+\ov K^0$
and $K^{*+}\ov K^0$ for $a_2=-0.51$, we obtain
\be  \label{FDeta}
F_0^{D\eta}(0)=0.446, \quad F_0^{D\eta'}(0)=0.287, \quad
F_0^{D_s\eta}(0)=-0.411, \quad F_0^{D_s\eta'}(0)=0.639. 
\en
Note that the form factor $F_0^{D_s\eta}$ has a sign opposite to 
$F_0^{D_s\eta'}$ due to the sign difference of the strange quark content
in the $\eta$ and $\eta'$ [see Eq.~(\ref{etapp})].
Using the above form factors for $D_s^+\to\etapp$ transition, 
we have computed the semileptonic decay rates of
$D_s^+\to\etapp e^+\nu$ and found an agreement with experiment.

   The form factors for $D\to\etapp$ and $D_s\to\etapp$ transitions also
have been calculated by BSW \cite{BSW} in a relativistic quark model. 
However, form factors obtained there did not include the wave function
normalizations and mixing angles.
\footnote{We are grateful to A. N. Kamal for pointing out this to us.}
For example, for $D\to\eta$ transition, BSW put in the $u\bar u$ constitutent
quark mass only, and for $D_s\to\eta$ the $s\bar s$ quark masses. 
In this way, BSW obtained \cite{BSW87}
\be
F_0^{D\eta_{u\bar u}}(0)=0.681, \quad F_0^{D\eta'_{u\bar u}}(0)=0.655, \quad
F_0^{D_s\eta_{s\bar s}}(0)=0.723, \quad F_0^{D_s\eta'_{s\bar s}}(0)=0.704.
\en
To compute the physical 
form factors one has to take into account the wave function
normalizations of $\eta$ and $\eta'$:
\be
F_0^{D\eta}=\left({1\over\sqrt{6}}\cos\theta-{1\over\sqrt{3}}\sin\theta\right)
F_0^{D\eta_{u\bar u}},  \qquad &&
F_0^{D\eta'}=\left({1\over\sqrt{6}}\sin\theta+{1\over\sqrt{3}}\cos\theta\right)
F_0^{D\eta'_{u\bar u}}, \non\\
F_0^{D_s\eta}=-\left({2\over\sqrt{6}}\cos\theta+{1\over\sqrt{3}}\sin\theta
\right)F_0^{D_s\eta_{s\bar s}},  \qquad &&
F_0^{D_s\eta'}=\left(-{2\over\sqrt{6}}\sin\theta+{1\over\sqrt{3}}\cos\theta
\right)F_0^{D_s\eta'_{s\bar s}}.
\en
Then the mixing angle $\theta=-10^\circ$ leads to 
\be  \label{bsw}
{\rm BSW}:~~~F_0^{D\eta}(0)=0.342, \quad F_0^{D\eta'}(0)=0.326, \quad
F_0^{D_s\eta}(0)=-0.509, \quad F_0^{D_s\eta'}(0)=0.500. 
\en
The above are the form factors used in the original BSW analysis for
$(D,D_s)\to(\eta,\eta')$ transitions \cite{BSW87}.
One can check that if $\theta=-22^\circ$ is used, the BSW form factors will be
close to ours as given in (\ref{FDeta}).

For the $q^2$ dependence of form factors in the region where $q^2$ is not 
too large, we shall use the pole dominance ansatz, namely,
$f(q^2)=f(0)/[1-(q^2/m^2_*)]^n$,
where $m_*$ is the pole mass given in \cite{BSW87}. 
A direct calculation of $D\to P$ and $D\to V$ 
form factors at time-like momentum transfer is available in the relativistic 
light-front quark model \cite{CCH} with the results that
the $q^2$ dependence of the form factors $A_0,~F_1$ is a dipole behavior 
(i.e. $n=2$), while $F_0$ exhibits a monopole dependence ($n=1$).
Note that in the BSW model, the $q^2$ dependence of $A_0,F_1$ is assumed 
to be the same as $F_0$, namely a monopole behaviour.

  Applying Eqs.~(\ref{a1a2}), (\ref{feta}), (\ref{FDeta})
and the form factor $A_0^{D\rho}(0)
=0.63$ \cite{CCH}, we have calculated the 
branching ratios for $(D^+,D_s^+)\to(\pi^+,\rho^+)\etapp$ and $D^0\to(\ov K^0,
\ov K^{0*})\etapp$ decays, as summarized in Table I (see the third
column), where use has been made of the charmed meson lifetimes \cite{PDG}
\be
\tau(D^0)=4.15\times 10^{-13}s, \quad \tau(D^+)=1.057\times 10^{-12}s, \quad 
\tau(D_s^+)=4.67\times 10^{-13}s.
\en
For comparison, the experimental measurements and the BSW predictions 
\cite{BSW87}
based on $a_1=1.25,~a_2=-0.51$, Eq.~(\ref{bsw}) for form factors 
$F_0^{D\etapp}$
and $F_0^{D_s\etapp}$ and a monopole $q^2$ dependence for all the form factors
are also exhibited in Table I. It is clear that our results differ from the
BSW predictions mainly for the decay modes $D^0\to\ov K^{*0}\eta,~D^+\to
\pi^+\eta$ and for the $\eta'/\eta$ ratio in $D_s^+\to\pi^+\etapp$ due
to the form factor differences in (\ref{FDeta}) and (\ref{bsw}) and the
$q^2$ dependence for form factors $A_0$ and $F_1$.
We see from Table I that the mixing angle $\theta=-22^\circ$ agrees better
with experiment than the angle $-10^\circ$ and 
that our predictions are in general 
consistent with experiment except for the decays: $D^0\to\ov K^0\eta',~D^+\to
\pi^+\eta$ and $D_s^+\to\rho^+\eta'$; the branching ratios of the first
two decay modes are too small by an order of magnitude. Hence, there are
three difficulties with the factorization approach in describing 
the hadronic $D$ decays to an $\eta$ and $\eta'$. First, it is naively 
expected
that $\b(D^0\to\ov K^0\eta')\ll\b(D^0\to\ov K^0\eta)$ due to the form 
factor suppression $F_0^{D\eta'}(0)/F_0^{D\eta}(0)=0.64$ and
the less phase space available to the former. However, experimentally
it is the other way around: $\b(D^0\to\ov K^0\eta')\sim 2.4\,\b(D^0\to\ov 
K^0\eta)$. Second, the predicted branching ratio for $D^+\to\pi^+\eta$
is too small by one order of magnitude. This is attributed to the
fact that the sign of $X_s^{(D\pi,\eta')}$ is opposite to $X_d^{(D\pi,
\eta')}$ and that there is a large cancellation between the external
$W$-emission amplitude $a_1X^{(D\eta,\pi)}$ and the internal $W$-emission
one $a_2(X_d^{(D\pi,\eta)}-X_s^{(D\pi,\eta)})$. Third, while the
generalized factorization is successful in predicting $\b(D_s^+\to\pi^+
\eta)$ and $\b(D_s^+\to\pi^+\eta')$ and marginally for $D_s^+\to 
\rho^+\eta$, its prediction for $\b(D_s^+\to\rho^+\eta')$ is too small
by about $2\sigma$ compared to experiment. This has motivated some authors
\cite{Ball} (see also \cite{Hoang}) to advocate an enhancement 
mechanism in which two gluons are produced in the $c\bar s$ annihilation
process and then hadronized into an $\eta'$.

\section{Final-state Interactions}
In the previous section we have 
pointed out three problems with the factorization approach for dealing with
the two-body $D$ decays to an $\eta$ or $\eta'$. One issue is that
final-state interactions (FSI) and nonspectator $W$-exchange or 
$W$-annihilation effects are not taken into account thus far. It is 
customary to argue that the $W$-exchange contribution is negligible due to
helicity and color suppression.
\footnote{In the factorization approach, the $W$-exchange amplitude in
$D\to P_1P_2$ decay is suppressed by a factor of $[(m_1^2-m_2^2)/m_D^2](
F_0^{P_1P_2}(m_D^2)/F_0^{DP_1}(m_2^2))$ relative to the external 
$W$-emission (assuming that $P_2$ is factored out). The form factor
$F_0^{P_1P_2}(q^2)$, which is antisymmetric in $P_1$ and $P_2$, is
suppressed at large momentum transfer $q^2=m_D^2$.}
Therefore, it is very unlikely that the nonspectator effects due to
$W$-exchange or $W$-annihilation can account for the large discrepancy 
between theory and experiment for $D^0\to\ov K^0\eta'$ and 
$D^+\to\pi^+\eta$. It remains to see if FSI could be the underlying 
mechanism responsible for the large enhancement of the
above-mentioned decay modes. The importance of FSI has long been 
realized in charm decay since some resonances are known to exist at 
energies close to the mass of the charmed meson. Consequently, the inelastic
scattering effects are crucial for understanding the pattern of charm weak 
decays. For example, the ratio $R=\Gamma(D^0\to\ov K^0\pi^0)/\Gamma(D^0\to 
K^-\pi^+)$ is predicted to be only of order $3\times 10^{-4}$ in the
naive factorization approach, while experimentally it is measured to be
$0.55\pm 0.06$ \cite{PDG}. It is known that the weak decay $D^0\to K^-\pi^+$
followed by the inelastic rescattering $K^-\pi^+\to\ov K^0\pi^0$ can raise
$\b(D^0\to\ov K^0\pi^0)$ dramatically and lower $\b(D^0\to K^-\pi^+)$ 
slightly.

   There are several different forms of FSI: elastic scattering and 
inelastic scattering such as quark exchange, resonance formation,$\cdots$, 
etc. As
emphasized in \cite{Zen}, the resonance formation of FSI via $q\bar q$
resonances is probably the most important one. Since FSI are 
nonperturbative in nature, in general it is notoriously difficult to calculate 
their effects. Nevertheless, as we shall see below, the effect of 
resonance-induced FSI can be
estimated provided that the mass and the width of the nearby resonances are
known. Before embarking on a detailed analysis, it is instructive to
elucidate qualitatively how resonant FSI work for the decay 
$D^0\to\ov K^0\eta'$ as an example.
Consider the weak decay $D^0\to K^-\pi^+$ followed by the strong-interaction
process: $K^-\pi^+\to{\rm scalar~resonances}\to\ov K^0\eta'$ (see Fig~1).
Note that Fig.~1 has the same topology as the $W$-exchange diagram, a point 
we will come back to later.
Denote the amplitude by $r_d$ ($r_s$) when the $d\bar d$ ($s\bar s$) 
pair is created and combined with the $s\bar d$ quarks to form the 
final state $\ov K^0\eta'$. Assuming SU(3) symmetry for the $d\bar d$ and
$s\bar s$ creation and taking the $\eta-\eta'$ mxing angle $\theta$ to be
$-19.5^\circ$, it is easily seen that
\be
A(D^0\to\ov K^0\eta')_{\rm FSI}=r_d+2r_s=3r_d, \quad A(D^0\to\ov K^0\eta)_{\rm
FSI}=r_d-r_s=0, ~ {\rm for}~\theta=-19.5^\circ.
\en
Since the branching ratio of $D^0\to K^-\pi^+$ is large enough, $\b(D^0
\to K^-\pi^+)=(3.83\pm 0.12)\%$ \cite{PDG}, it is quite plausible that
resonance-induced FSI could enhance $\b(D^0\to\ov K^0\eta')$ 
by an order of magnitude without affecting the original good
agreement between theory and experiment for $D^0\to \ov K^0\eta$. Therefore,
this mechanism enables us to understand why the decay rate of
$D^0\to\ov K^0\eta'$ is larger than $D^0\to\ov K^0\eta$, even though the
factorizable contribution to the former is smaller than the latter.

\begin{figure}[ht]
\vspace{1cm}
\hskip 3.8cm
  \psfig{figure=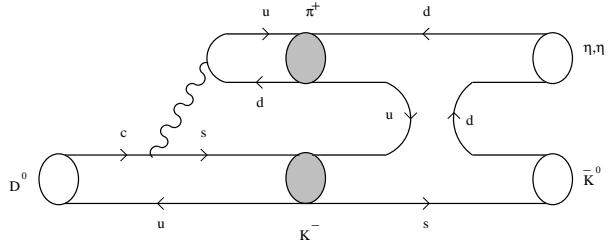,width=8cm}
\vspace{0cm}
    \caption[]{\small Contributions to $D^0\to\ov K^0\eta(\eta')$ from the
weak decay $D^0\to K^-\pi^+$ followed by a resonant rescattering.}
\end{figure}

    We will repeat the analysis of \cite{Zen} to study the effects of 
resonant FSI for the decays $D^0\to\ov K^0\etapp$. It turns out that 
the quark-diagram approach put forward in \cite{CC87,CC86} is quite suitable 
for this purpose. In this approach, all two-body nonleptonic weak 
decays of charmed mesons can be expressed in terms of six distinct
quark diagrams: $\a$, the external $W$-emission diagram; $\b$, the internal
$W$-emission diagram; $\c$, the $W$-exchange diagram; $\d$, the 
$W$-annihilation diagram; ${\cal E}$, the horizontal $W$-loop diagram; and 
${\cal F}$, the vertical $W$-loop diagram.
It should be stressed that these quark diagrams are classified according 
to their topologies and hence they are {\it not} Feynman
graphs. The quark-diagram amplitudes for $D^0\to K^-\pi^+,~\ov 
K^0\pi^0,~\ov K^0\eta_{ns}$ and $\ov K^0\eta_s$, where $\eta_{ns}={1\over
\sqrt{2}}(u\bar u+d\bar d)$ and $\eta_s=s\bar s$, are given by (see 
Table III of \cite{CC87}):
\be \label{qds}
&& A(D^0\to(\ov K\pi)_{3/2}) = {1\over\sqrt{3}}(\a+\b),   \qquad
A(D^0\to(\ov K\pi)_{1/2}) = {1\over\sqrt{6}}(2\a-\b+3\c),   \non \\
&& A(D^0\to\ov K^0\eta_{ns}) = {1\over\sqrt{2}}(\b+\c), \qquad
A(D^0\to\ov K^0\eta_s) = \c.
\en
For FSI through $q\bar q$ resonances, we consider the $D$-type coupling 
for the strong interaction $P_1P_2\to P'$ ($P'$: scalar
meson), namely $\kappa{\rm Tr}\left(P'\{P_1,\,P_2\}\right)$ with $\kappa$
being a flavor-symmetric strong coupling \cite{Zen}. Noting that
$(\ov K\pi)_{3/2}$ does not couple to $(\ov K\pi)_{1/2}$, $\ov K^0\eta_{ns}$,
and $\ov K^0\eta_s$ via FSI, the strong reaction matrix $K_0$, which is
related to the $S$ matrix by $S_0=(1+iK_0)/(1-iK_0)$, for the 
$I={1\over 2}$ sector has the form:
\be
K_0=\kappa^2\left(\matrix{ {3\over 2} & {\sqrt{3}\over 2} &
{\sqrt{3}\over\sqrt{2}} \cr   {\sqrt{3}\over 2} & {1\over 2} & {1\over
\sqrt{2}} \cr    {\sqrt{3}\over\sqrt{2}}  & {1\over\sqrt{2}} & 1    \cr}
\right)
\en
in the basis of $(\ov K\pi)_{1/2},~\ov K^0\eta_{ns},~\ov K^0\eta_s$.
The eigenvalues and eigenvectors of the $K_0$ matrix are
\be  \label{eigen}
\lambda_1=3\kappa^2, \quad  && (PP)_1={1\over\sqrt{6}}\left[ \sqrt{3}
(\ov K\pi)_{1/2}+\ov K^0\eta_{ns}+\sqrt{2}(\ov K^0\eta_s)\right],   \non\\
\lambda_2=0, \quad  && (PP)_2={1\over\sqrt{6}}\left[- \sqrt{3}
(\ov K\pi)_{1/2}+\ov K^0\eta_{ns}+\sqrt{2}(\ov K^0\eta_s)\right],   \non\\
\lambda_3=0, \quad  && (PP)_3={1\over\sqrt{3}}\left[ \sqrt{2}
(\ov K^0\eta_{ns})-\ov K^0\eta_s\right].
\en
In this new basis, the weak decay amplitudes are unitarized by FSI as
\cite{Zen}
\be   \label{Watson}
A\left(D^0\to (PP)_i\right)\to \cos\delta_i e^{i\delta_i}A\left( D^0\to 
(PP)_i\right),
\en
as required by the unitarity of the $S$ matrix (known as Watson's theorem)
with $\delta_i$ being the eigenphases of the $K$ matrix. It is
then straightforward to show from Eqs.~(\ref{qds},\ref{eigen},\ref{Watson}) 
that resonance-induced FSI
amount to modifying the $W$-exchange amplitude by \cite{Zen}
\be \label{newc}
\c\to \c+(\c+{1\over 3}\a)(\cos\delta_1 e^{i\delta_1}-1)
\en
and leaving the other quark-diagram amplitudes intact, where 
$\delta_1=3\kappa^2$. This is 
consistent with what has been expected before: The resonance 
contribution to FSI, which arises mainly from the external $W$-emission 
diagram for the decay $D^0\to (\ov K\pi)_{1/2}$ followed by final-state 
$q\bar q$ resonance, has the same topology as the $W$-exchange quark diagram.
We thus see that even if the short-distance $W$-exchange vanishes, as 
commonly asserted, a long-distance $W$-exchange still can be induced via
FSI in resonance formation.

   Substituting (\ref{newc}) back into (\ref{qds}) and neglecting the 
short-distance $W$-exchange contribution, we obtain
\be
 A(D^0\to\ov K^0\eta) = {G_F\over\sqrt{2}}V_{cs}^*
V_{ud}\left[a_2X^{(D\eta,K)}- a_1X^{(DK,\pi)}\,{\cos\delta 
e^{i\delta}-1\over 3}\left(
{\cos\theta\over\sqrt{6}}+{2\over\sqrt{3}}\sin\theta\right)\right],  \non\\
 A(D^0\to\ov K^0\eta') = {G_F\over\sqrt{2}}V_{cs}^*
V_{ud}\left[a_2X^{(D\eta',K)}- a_1X^{(DK,\pi)}\,{\cos\delta 
e^{i\delta}-1\over 3}\left(
{\sin\theta\over\sqrt{6}}-{2\over\sqrt{3}}\cos\theta\right)\right], \non\\
\en
where $a_1X^{(DK,\pi)}$ is the factorizable amplitude for $D^0\to K^-\pi^+$ 
and $X^{(DK,\pi)}=if_\pi(m_D^2-m_K^2)F_0^{DK}(m_\pi^2)$.

  In order to determine the phase shift $\delta$, we shall assume that 
there exist nearby resonances in the charmed-meson mass region and that 
the phase is related to the Breit-Wigner resonance by
\be
{1\over 2i}(e^{2i\delta}-1)=\sin\delta\,e^{i\delta}=\,{\Gamma_*\over
2(m_*-m_D)-i\Gamma_*}\,,
\en
in the rest frame of the charmed meson,
where $m_*$ and $\Gamma_*$ are the mass and width of the 
resonance, respectively. It follows that
\be
\tan\delta={\Gamma_*\over 2(m_*-m_D)} \,.
\en
For parity-violating $D\to PP$ decays, there is a $0^+$ resonance 
$K^*_0(1950)$ in $(s\bar d)$ quark content with mass $1945\pm 10\pm 20$ MeV 
and width $210\pm 34\pm 79$ MeV \cite{PDG}. It is clear from Table I that the
resultant branching ratio of $D^0\to\ov K^0\eta'$ is 
enhanced by resonance-induced
FSI by one order of magnitude, whereas $D^0\to\ov K^0\eta$ remains
essentially unaffected. Therefore, we conclude that it is the final-state
interaction that accounts for the bulk of $\b(D^0\to\ov K^0\eta')$ 
and explains its larger decay rate than $D^0\to\ov K^0\eta$.

   For decays $D^0\to\ov K^{*0}\etapp$, they can proceed through the processes
$D^0\to K^{*-}\pi^+,K^-\rho^+\to\ov K^{*0}\etapp$. Following the 
quark-diagram notation of \cite{CC87} that primed amplitudes are for 
the case that the vector meson is produced from the charmed quark decay,
we write
\be
A(D^0\to\ov K^{*0}\eta_{ns})={1\over\sqrt{2}}(\b'+\c'),  \qquad
A(D^0\to\ov K^{*0}\eta_s)=\c.  
\en
Repeating the same analysis as before, one obtains (see \cite{Zen} 
for details)
\be
&& \c \to \c+{1\over 2}\left[\c+\c'+{1\over 3}(\a+\a')\right](\cos\delta
e^{i\delta}-1),   \non \\
&& \c' \to \c'+{1\over 2}\left[\c+\c'+{1\over 3}(\a+\a')\right](\cos\delta
e^{i\delta}-1),   
\en
where $\a$ is the external $W$-emission amplitude for $D^0\to K^-\rho^+$ and
$\a'$ for $D^0\to K^{*-}\pi^+$. Neglecting the short-distance 
$W$-exchange, we obtain
\be
A(D^0\to \ov K^{*0}\eta) &=& {G_F\over\sqrt{2}}V_{cs}^*V_{ud}\Bigg[ a_2X^{(
D\eta,K^*)}   \non \\
&-& a_1\left(X^{(DK^*,\pi)}+X^{(DK,\rho)}\right)\,{\cos\delta e^{i\delta}-1
\over 6}\left({\cos\theta\over\sqrt{6}}+{2\over\sqrt{3}}\sin\theta\right)
\Bigg],   \non \\
A(D^0\to \ov K^{*0}\eta') &=& {G_F\over\sqrt{2}}V_{cs}^*V_{ud}\Bigg[ a_2X^{(
D\eta',K^*)}   \non \\
&-& a_1\left(X^{(DK^*,\pi)}+X^{(DK,\rho)}\right)\,{\cos\delta e^{i\delta}-1
\over 6}\left({\sin\theta\over\sqrt{6}}-{2\over\sqrt{3}}\cos\theta\right)
\Bigg], 
\en
with $X^{(DK^*,\pi)}=-2f_\pi m_{K^*}A_0^{DK^*}(m_\pi^2)(\vp\cdot p_{_D})$ 
and $X^{(DK,\rho)}=-2f_\rho m_\rho F_1^{DK}(m_\rho^2)(\vp\cdot p_{_D})$. 
The relevant $0^-$ resonance for
$D\to \ov K^*\etapp$ decays is the $K(1830)$ with mass $\sim$ 1830 MeV and
width $\sim$ 250 MeV \cite{PDG}. As shown in Table I, the resonance 
effect has almost no 
impact on $D^0\to\ov K^{*0}\eta$. The smallness of $\b(D^0\to\ov 
K^{*0}\eta')$ of order $2\times 10^{-4}$ is due mainly to the 
severe phase-space suppression. Note that our predictions for 
$D^0\to\ov K^0\eta'$ and $\ov K^{*0}\eta$ are still slightly smaller than
experiment and that so far we have not considered the effects of
$W$-exchange and FSI other than resonance formation.

We next turn to the Cabibbo-suppressed decays $D^+\to(\pi^+,\rho^+)\etapp$.
As noted in passing, in the absence of FSI, the branching ratio of $D^+\to
\pi^+\eta$ is very small, of order $10^{-4}$, owing to a large cancellation
between external and internal $W$-emission amplitudes. Since $D^+\to K^+
\ov K^0$ has a relatively large branching ratio,
$\b(D^+\to K^+\ov K^0)=(7.2\pm 1.2)\times 10^{-3}$ \cite{PDG},
it is conceivable that $D^+\to\pi^+\eta$ can receive significant 
contributions from resonant FSI through the process $D^+\to K^+\ov K^0
\to\pi^+\eta$. (Note that $\pi^+\pi^0$ does not couple to $\pi^+\etapp$ 
by strong interactions.) The quark diagram amplitudes for $D^+\to\pi^+\etapp$
are given by \cite{CC87}
\be
&& A(D^+\to K^+\ov K^0)=-(\a-\d), \qquad
A(D^+\to\pi^+\eta_{ns})={1\over\sqrt{2}}(\a+\b+2\d),  \non \\
&& A(D^+\to\pi^+\eta_s)=-\b.
\en
Proceeding as before, resonance-induced coupled-channel
effects among the three channels: $K^+\ov K^0,\,\pi^+\eta_{ns}$ and 
$\pi^+\eta_s$
will only modify the magnitude and phase of the $W$-annihilation 
amplitude and leave the other quark-diagram amplitudes unaffected:
\be
\d\to \d+\left(\d+{1\over 3}\a\right)(\cos\delta e^{i\delta}-1).
\en
Hence,
\be
A(D^+\to\pi^+\eta) &=& {G_F\over\sqrt{2}}V_{cd}^*V_{ud}\Bigg[ a_1X^{(D\eta,
\pi)}+a_2\left(X_d^{(D\pi,\eta)}-X_s^{(D\pi,\eta)}\right)   \non \\
&-& {\sqrt{2}\over 3}a_1X^{(DK,K)}(\cos\delta
e^{i\delta}-1)\left({\cos\theta\over\sqrt{3}}-\sqrt{2\over 3}\sin\theta
\right)\Bigg],   \non \\
A(D^+\to\pi^+\eta') &=& {G_F\over\sqrt{2}}V_{cd}^*V_{ud}\Bigg[ a_1X^{(D\eta',
\pi)}+a_2\left(X_d^{(D\pi,\eta')}-X_s^{(D\pi,\eta')}\right)    \non \\ 
&-& {\sqrt{2}\over 3}a_1X^{(DK,K)}(\cos\delta
e^{i\delta}-1)\left({\sin\theta\over\sqrt{3}}+\sqrt{2\over 3}\cos\theta
\right)\Bigg],   
\en
where $X^{(DK,K)}=if_K(m_D^2-m_K^2)F_0^{DK}(m_K^2)$.

 A nearby $0^+$ resonance $a_0$ in the charm mass region has not 
been observed. We shall follow \cite{Buccella}
to employ $m_{a_0}=1869$ MeV and $\Gamma_{a_0}=300$ MeV, where the mass is
estimated from the equispacing formula $m^2_{a_0}=m^2_{K^*_0}-m^2_K-m^2_\pi$.
Numerically, both $\b(D^+\to\pi^+\eta)$ and $\b(D^+\to\pi^+\eta')$ are
enhanced, in particular the former is increased by 
an order of magnitude (see Table I). 

   Contrary to $\pi^+\eta$ and $\pi^+\eta'$ final states, resonant FSI are
negligible for $\rho^+(\eta,\eta')$ states for the following reason. The
$G$ parity of $\rho\eta$ and $\rho\eta'$ is even, while the $J=0,~I=1$
meson resonance made from a quark-antiquark pair (i.e. $u\bar d$)
has odd $G$ parity. This is also true for the $W$-annihilation 
process $c\bar d\to u\bar d$. As stressed in \cite{Lipkin}, the even--$G$ 
state $\rho\eta$ or $\rho\eta'$ does not couple to any single meson 
resonances, nor to the state produced by the $W$-annihilation diagram 
with no gluons emitted by the initial state before annihilation.
We would like to remark that at the factorizable
amplitude level $|A(D^+\to\rho^+\eta')|>|A(D^+\to\rho^+\eta)|$, but
$\b(D^+\to\rho^+\eta')<\b(D^+\to\rho^+\eta)$ due to the lack of phase space
available to the former.

   As for $D_s^+\to\pi^+\etapp$ decays, the quark diagram amplitudes are
\be \label{ds}
A(D_s^+\to K^+\ov K^0)=\b+\d, \quad
A(D_s^+\to\pi^+\eta_{ns})=\sqrt{2}\,\d, \quad A(D_s^+\to\pi^+\eta_s)=\a.
\en
The analysis of resonant coupled-channel effects is the same as $D^+\to\pi^+
\etapp$ and it leads to \cite{Zen}
\be \label{newd}
\d\to \d+\left(\d+{1\over 3}\b\right)(\cos\delta e^{i\delta}-1),
\en
where $\b$ is the internal $W$-emission amplitude for $D_s^+\to K^+\ov K^0$.
Neglecting $W$-annihilation as before, we obtain from (\ref{ds}) and 
(\ref{newd}) that
\be
A(D_s^+\to\pi^+\eta) = {G_F\over\sqrt{2}}V_{cs}^*V_{ud}\Bigg[a_1X^{(D_s\eta,
\pi)}+{\sqrt{2}\over 3}a_2X^{(D_sK,K)}(\cos\delta e^{i\delta}-1)\Big({\cos
\theta\over\sqrt{3}}-\sqrt{2\over 3}\sin\theta\Big)\Bigg],   \non \\
A(D_s^+\to\pi^+\eta') = {G_F\over\sqrt{2}}V_{cs}^*V_{ud}\Bigg[a_1X^{(D_s
\eta',\pi)}+{\sqrt{2}\over 3}a_2X^{(D_sK,K)}(\cos\delta e^{i\delta}-1)\Big({
\sin\theta\over\sqrt{3}}+\sqrt{2\over 3}\cos\theta\Big)\Bigg], \non\\
\en
with $X^{(D_s K,K)}=if_K(m^2_{D_s}-m^2_K)F_0^{D_sK}(m_K^2)$.
It is interesting to remark that $D_s^+\to\pi^+\eta$ is suppressed in the
presence of resonant FSI, while $D_s^+\to\pi^+\eta'$ is enhanced (see Table 
I). This is ascribed to the fact that the external $W$-emission amplitudes for
$D_s^+\to\pi^+\eta$ and $\pi^+\eta'$ are opposite in sign due to a 
relative sign 
difference between the form factors $F_0^{D_s\eta}$ and $F_0^{D_s\eta'}$.

The same argument that resonance-induced FSI and $W$-annihilation without gluon
emission in the initial state do not contribute to $D^+\to\rho^+\etapp$
also applies to $D_s^+\to\rho^+\etapp$. As a consequence, the large observed
branching ratio of $D_s^+\to\rho^+\eta'$ is surprising.
Theoretically, it is very difficult to raise the branching ratio of
the $\rho\eta'$ mode from 3.9\% to the level of 10\% without suppressing 
$D_s^+\to\rho^+\eta$. First, in general the effect of FSI is useful 
and significant for the weak decay $D\to X$ only if there exists a decay
$D\to Y$ with a sufficiently large decay rate, i.e. $\b(D\to Y)\gg\b(D\to X)$,
and if $X$ and $Y$ channels 
couple through FSI. For $D_s^+\to VP$ decays, the branching 
ratio of $D_s^+\to\phi\pi^+$ 
is only 3.6\% \cite{PDG}, which is even smaller than $D_s^+\to\rho^+\eta$.
Hence, FSI in any form are unlikely to raise 
$\b(D_s^+\to\rho^+\eta',\rho^+\eta)$ substantially. Second,
an enhancement mechanism has been suggested in \cite{Ball} that a $c\bar s$ 
pair annihilates into a $W^+$ and two gluons, then the two gluons 
hadronize into a favor-singlet $\eta_0$. Since 
$\eta_0=\eta'\cos\theta-\eta\sin\theta$ and
the mixing angle $\theta$ is negative, it is evident that if $\b(D_s^+
\to\rho^+\eta')$ is enhanced by this mechanism, $\b(D_s^+\to\rho^+\eta)$ 
will be suppressed due to the destructive interference between the 
external $W$-emission and the gluon-mediated process, recalling that the
external $W$-emission amplitudes for $D_s^+\to\rho^+\eta$ and $D_s^+
\to\rho^+\eta'$ are opposite in sign. Hence, if $\b(D_s^+\to\rho^+\eta')$
is accommodated by this new mechanism, then we will have a hard time to
explain $\b(D_s^+\to\rho^+\eta)$. The $W$-annihilation diagram, which is 
not subject to color and helicity suppression
in $(D_s^+,D^+)\to\rho^+\etapp$ decays, is expected to play some role. Even 
a small contribution from $W$-annihilation, say $\d/\a\sim 0.2$, can easily
increase the decay rate by a factor of 2. However, by the same reasoning 
as shown above, when $W$-annihilation raises the branching ratio of one of 
the $D_s^+\to\rho^+\etapp$ decay modes, it will lower the other one.
Third, the phase-space factor relevant to
$D_s^+\to\rho^+\etapp$ is $p_c[(m_{D_s}^2-m_\rho^2-m_\etapp^2)^2-4m_\rho^2
m_\etapp^2]$ with $p_c$ being the c.m. momentum. The phase-space suppression
of $\rho\eta'$ relative to $\rho\eta$ is found to be 0.27. In order to 
achieve $\b(D_s^+\to\rho^+\eta')\sim\b(D_s^+\to\rho^+\eta)\sim 10\%$,
a new mechanism must be introduced to overcome the phase-space 
suppression for the former and
in the meantime it should not lower the decay rate of the latter. 
To our knowledge, it is difficult to speculate such a mechanism.

   Since the decay rates of $D_s^+\to\rho^+\etapp$ are sensitive to the
form factors $F_1^{D_s\etapp}$, it is advantageous to consider the ratios
$R_\etapp\equiv\Gamma(D_s^+\to\rho^+\etapp)/\Gamma(D_s^+\to\etapp
e^+\nu)$ in order to test the generalized factorization hypothesis. 
Neglecting $W$-annihilation, factorization leads to the 
form-factor-independent predictions $R_\eta=2.9$ and $R_{\eta'}=3.5$,
while experimentally $R_\eta=4.4\pm 1.2$ and $R_{\eta'}
=12.0\pm 4.3$ \cite{CLEO}. (Our value for $R_{\eta'}$ is slightly different
from the result $R_{\eta'}=2.9$ obtained in \cite{Kamal} as we use a
dipole $q^2$ dependence for the form factors $F_1^{D_s\etapp}(q^2)$.)
We have argued that FSI, $W$-annihilation and 
the production of excess $\eta'$ from gluons are not
helpful in understanding the very large branching ratios of
$D_s^+\to \rho^+\etapp$. Hence, the very large value of $R_{\eta'}$ remains
as an enigma.

\section{Conclusions}
With the improved $(D,D_s^+)\to(\eta,\eta')$
form factors and decay constants of the $\eta$ and $\eta'$, we have 
employed the 
generalized factorization approach to reanalyze the decays of charmed mesons
into the final states containing an $\eta$ or $\eta'$. We show that 
resonant FSI
are able to enhance $\b(D^0\to\ov K^0\eta')$ and $\b(D^+\to\pi^+\eta)$ by
an order of magnitude. Resonance-induced couple-channel effects will
suppress $D_s^+\to\pi^+\eta$ and enhance $D_s^+\to\pi^+\eta'$. Contrary
to $D\to P\etapp$ decays, resonant FSI play only a minor role for
$D^0\to\ov K^{*0}\etapp$ and do not contribute to $(D^+,D_s^+)\to\rho^+
\etapp$. We argue that it is difficult to understand the 
observed large decay rates of the $\rho^+\eta'$ and $\rho^+\eta$ decay modes 
of $D_s^+$ simultaneously. FSI are not helpful due to the
absence of $D_s^+\to PP$ decays that have much larger decay rates than 
$D_s^+\to\rho^+\eta'$. $W$-annihilation and 
a possible production of the $\eta'$ due to gluon-mediated processes
can in principle enhance $\b(D_s^+\to\rho^+\eta')$, but, unfortunately, 
they will also suppress $\b(D_s^+\to\rho^+\eta)$.

\vskip 1.5 cm
\acknowledgments
We are grateful to A. N. Kamal for a helpful discussion on te BSW for factors.
This work was supported in part by the National 
Science Council of ROC under Contract Nos. NSC87-2112-M-001-048 and
NSC87-2112-M-006-018.

\renewcommand{\baselinestretch}{1.1}
\newcommand{\bi}{\bibitem}
%

\newpage
\vskip 0.1cm
\begin{table}[ht]
{\small Table I. Branching ratios (in units of \%) of the charmed meson 
decays to an $\eta$ or $\eta'$. The BSW predictions \cite{BSW87} are for the 
$\eta-\eta'$ mixing angle $\theta=-10^\circ$, while ours are for $\theta=-
22^\circ$.}
\begin{center}
\begin{tabular}{ l c c c c } 
 &  & \multicolumn{2}{c}{This work} & \\\cline{3-4} 
\raisebox{1.5ex}[0cm][0cm]{Decay} & \raisebox{1.5ex}[0cm][0cm]{BSW
\cite{BSW87}} & without FSI & with resonant FSI & \raisebox{1.5ex}[0cm]
[0cm]{Expt. \cite{CLEO,PDG} } \\ 
\hline
$D^0\to\ov K^0\eta$ & 0.31 & 0.50 & $0.54^{+0.01}_{-0.02}$ & $0.71\pm 0.10$ \\
$D^0\to\ov K^0\eta'$ & 0.12 & 0.10 & $0.90^{+0.27}_{-0.45}$ & $1.72\pm 0.26$\\
$D^0\to\ov K^{*0}\eta$ & 0.28 & 0.76 & 0.74 & $1.9\pm 0.5$   \\
$D^0\to\ov K^{*0}\eta'$ & 0.002 & 0.004 & 0.02 & $<0.11$   \\
\hline
$D^+\to\pi^+\eta$ & 0.002 & 0.011 & 0.12 & $0.30\pm 0.06$  \\
$D^+\to\pi^+\eta'$ & 0.15 & 0.25 & 0.59 & $0.50\pm 0.10$  \\
$D^+\to\rho^+\eta$ & 0.06 & 0.20 & 0.20 & $<0.68$  \\
$D^+\to\rho^+\eta'$ & 0.03 & 0.07 & 0.07 & $<0.52$  \\
\hline
$D_s^+\to\pi^+\eta$ & 3.66 & 2.43 & 1.30 & $1.73\pm 0.47$ \\
$D_s^+\to\pi^+\eta'$ & 2.14 & 3.32 & 4.37 & $3.71\pm 0.98$ \\
$D_s^+\to\rho^+\eta$ & 6.87 & 5.92 & 5.92 & $10.7\pm 3.1$ \\
$D_s^+\to\rho^+\eta'$ & 1.94 & 3.86 & 3.86 & $10.0\pm 2.9$ \\
\end{tabular}
\end{center}  
\end{table}


\begin{thebibliography}{99}
%

\bi{CT98} H.Y. Cheng and B. Tseng, hep-ph/9803457; H.Y. Cheng, talk 
presented at the First APCTP Workshop on
Pacific Particle Physics Phenomenology, Seoul, Oct. 31-Nov. 2, 1997 
[hep-ph/9712244].

\bi{CLEO} CLEO Collaboration, C.P. Jessop {\it et al.,} CLNS 97/1515 
[hep-ex/9801010].

\bibitem{BSW87} M. Bauer, B. Stech, and M. Wirbel, \zp {\bf C34}, 103 (1987).

\bi{Verma} R.C. Verma and A.N. Kamal, \pr {\bf D43}, 829 (1991); 
A.N. Kamal, Q.P. Xu, and A. Czarnecki, \pr {\bf D48}, 5215 (1993); 
R.C. Verma, A.N. Kamal, and M.P. Khanna, \zp {\bf C65}, 255 (1995);
K.K. Sharma, A.C. Katoch, and R.C. Verma, \zp {\bf C75}, 253 (1997).

\bi{Kamal} A.N. Kamal, Q.P. Xu, and A. Czarnecki, \pr {\bf D49}, 1330 (1994). 

\bi{Lipkin92} H.J. Lipkin, \pl {\bf B283}, 421 (1992); P. Bedaque, A. Das, 
and V.S. Mathur, \pr {\bf D49}, 269 (1994); I. Hinchliffe and T.A. Kaeding, 
\pr {\bf D54}, 914 (1996).

\bi{Buccella} F. Buccella, M. Lusignoli, and A. Pugliese, \pl {\bf B379}, 249
(1996); F. Buccella, M. Lusignoli, G. Miele, A. Pugliese, and P. Santorelli,
\pr {\bf D51}, 3478 (1995).


\bi{Cheng94} H.Y. Cheng, \pl {\bf B395}, 345 (1994); in {\it Particle Theory
and Phenomenology,}  XVII International Karimierz Meeting on Particle
Physics, Iowa State University, May 1995, edited by K.E. Lassila {\it et al}.
(World Scientific, Singapore, 1996), p.122 [hep-ph/9506340].

\bi{BSW} M. Wirbel, B. Stech, and M. Bauer, \zp {\bf C29}, 637 (1985).

\bi{Chau1} L.L. Chau, H.Y. Cheng, W.K. Sze, H. Yao, and B. Tseng, \pr
{\bf D43}, 2176 (1991).

\bi{Holstein} E.P. Venugopal and B.R. Holstein, \pr {\bf D57}, 4397 (1998).

\bi{Dpi} CLEO Collaboration, F. Butler {\it et al}., \pr {\bf D52}, 2656
(1995); J. Bartelt {\it et al.,} \pl {\bf B405}, 373 (1997).

\bi{CCH} H.Y. Cheng, C.Y. Cheung, and C.W. Hwang, \pr {\bf D55}, 1559 (1997).

\bi{PDG} C. Caso {\it et al.} (Particle Data Group), {\sl Eur. Phys. J.}
{\bf C3}, 1 (1998).

\bi{Ball} P. Ball, J.-M. Fr\`ere, and M. Tytgat, \pl {\bf B365}, 367 (1996).

\bi{Hoang} N.L. Hoang, A.V. Nguyen, and X.Y. Pham, \pl {\bf B357}, 177 (1995).

\bi{Zen} P. \.Zenczykowski, {\sl Acta Phys. Polon.} {\bf B28}, 1605 (1997) 
[hep-ph/9601265].

\bi{CC87} L.L. Chau and H.Y. Cheng, \pr {\bf D36}, 137 (1987).

\bi{CC86} L.L. Chau and H.Y. Cheng, \prl {\bf 56}, 1655 (1986); \pl {\bf
B222}, 285 (1989).

\bi{Lipkin} H.J. Lipkin, hep-ph/9708253.

\end{thebibliography}
\end{document}